\documentclass[twocolumn, pra, superscriptaddress,notitlepage]{revtex4-1}
\usepackage{graphicx}
\usepackage{physics}
\usepackage{epsfig}
\usepackage{color}
\usepackage{xspace}
\usepackage{array}
\usepackage{amsmath}
\usepackage{amsfonts}
\usepackage{ulem}
\usepackage[dvipsnames]{xcolor}
\usepackage[breaklinks,colorlinks,linkcolor=blue,citecolor=blue,urlcolor=blue]{hyperref}

\begin{document}
\title{Analogue Black Holes in Reactive Molecules}

\author{Ren Zhang}
\thanks{They contribute equally to this work.}
\affiliation{School of Physics, Xi'an Jiaotong University, Xi'an, Shaanxi 710049, China }
\author{Chenwei Lv}
\thanks{They contribute equally to this work.}
\affiliation{Department of Physics and Astronomy, Purdue University, West Lafayette, IN, 47907, USA}
\author{Qi Zhou}
\email{zhou753@purdue.edu}
\affiliation{Department of Physics and Astronomy, Purdue University, West Lafayette, IN, 47907, USA}
\affiliation{Purdue Quantum Science and Engineering Institute, Purdue University, West Lafayette, IN, 47907, USA}

\date{\today}

\begin{abstract}
We show that reactive molecules with a unit probability of reaction naturally provide a simulator of some intriguing black hole physics. 
The unit reaction at the short distance acts as an event horizon and delivers a one-way traffic for matter waves passing through the potential barrier when two molecules interact by high partial-wave scatterings or dipole-dipole interactions.  
In particular, the scattering rate as a function of the incident energy exhibits a thermal-like distribution near the maximum of the interaction energy in the same manner as a scalar field scatters with the potential barrier outside the event horizon of a black hole. Such a thermal-like scattering can be extracted from the temperature-dependent two-body loss rate measured in experiments on KRb and other molecules. 
\end{abstract}

\maketitle

\section{Introduction}
Black holes (BHs) give rise to a variety of intriguing phenomena in our universe. 
A famous example is Hawking radiation that produces quantum particles emitted from the event horizon~\cite{Hawkingradiation} .  Another important problem in BH physics concerns a different scenario that a matter or gravitational wave in the exterior of BH travels towards it and scatters with a potential barrier produced by the BH spacetime metric outside the event horizon~\cite{Susskind}.
%The event horizon provides a perfect absorption for any objects traveling towards it. As such, once a matter or gravitational wave passes through a barrier produced by the black hole spacetime outside the event horizon, it must propagate in a one-way traffic without returning. The potential barrier itself is responsible for producing quasi-normal modes and BH ringdown 
Since the event horizon provides a perfect absorption for any objects traveling towards it,  once a matter or gravitational wave passes through the potential barrier, it must propagate in a one-way traffic without returning. The potential barrier itself is responsible for producing quasinormal modes and BH ringdown
\cite{QNM,QNM2,QNM3,QNM4}. Near its maximum, the potential barrier can be well approximated by an inverted parabola. The resultant transmission and reflection rate exhibit thermal-like behaviors similar to a quantum mechanical problem of a particle scattered with an inverted harmonic potential \cite{landaubook,Chandrasekhar1998,Iyer1987-1}. The effective temperature in this thermal-like scattering encodes the mass of a Schwarzschild black hole. Interestingly, the inverted harmonic oscillator (IHO) also underlies the profound Hawking-Unruh radiation~\cite{Unruh0,book,IHO2,Smitha,IHO_OTOC}.
Both the dynamics near the event horizon and in the accelerating reference frame can be mapped to IHOs.

While significant progress has been made in observational astronomy in the study of BHs in the past few decades \cite{GW1,GW2}, there have been long-term efforts of exploring analogue BHs in laboratories. Therein, the spacetime metrics of BHs can be synthesized using a variety of platforms, such as water or supersonic fluid~\cite{Unruh,Unruh2,Gregory,Renaud,Germain,Silke,Visser1998,Giovanazzi2005,Rousseaux2008}, Bose-Einstein condensates~\cite{Jeff,Jeff2,Jeff3,Zoller2000,Carusotto2008}, %metamaterials
artificial optical materials
~\cite{Leonhardt,Leonhardt2,Genov}, and superconductor circuits~\cite{Fanheng}. In parallel, it has been found that certain quantum systems can be used to simulate scatterings problems of BHs. For instance, when quantum Hall states are subject to saddle potentials, an effective inverted harmonic potential arises, and the transmission and reflection rate become thermal-like \cite{Stone,Smitha,Fertig}. Though such a deep connection between transport phenomena of quantum Hall states and BH physics has attracted long-lasting theoretical interest, it eludes experiments on two-dimensional electron gases.

\begin{figure}
\centering
\includegraphics[width=0.4\textwidth]{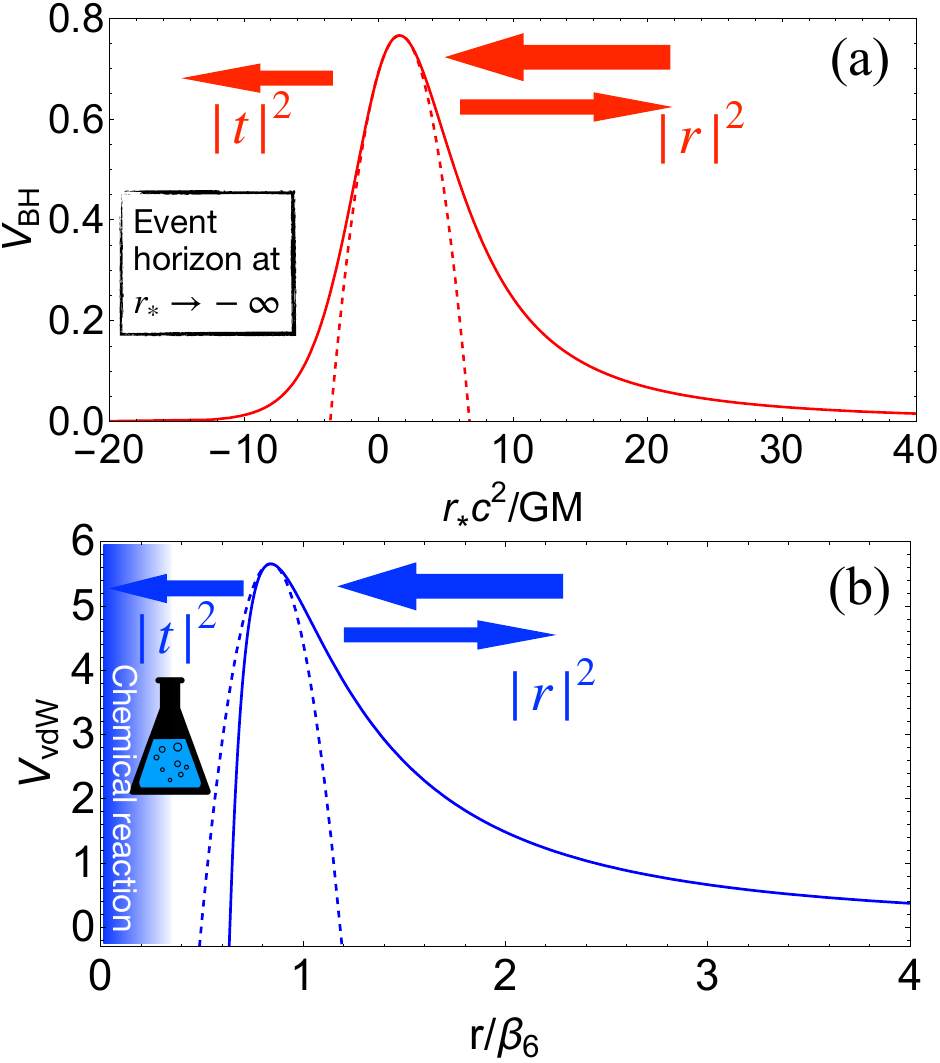} 
\caption{(a): The potential barrier (solid curves) outside the event horizon of the Schwarzschild black hole. (b): The interaction between two molecules with high partial-wave scattering or dipole-dipole interactions. At short distance, the unit probability of reaction leads to an absorbing boundary condition that mimics the event horizon. In both (a) and (b), $|r|^{2}$ and $|t|^{2}$ denote the reflection and transmission rate, respectively, and dashed curves denote the IHO approximation.\label{Fig1}}
\end{figure}

In this work, we point out that reactive molecules with a unit probability of reaction at a short distance provide a natural simulator to study scattering problems in BHs.
While the long-range part of the interaction between two molecules is a van der Waals potential, chemical reactions occur or two molecules form long-lived complexes in the short range~\cite{complex0,complex,complex1}. 
Some molecules like KRb have a unit probability of reaction such that whenever the separation between two molecules decreases down to a certain short length, chemical reactions occur with 100\% probability~\cite{Yejun0,Yejun3}. 
In other words, the incident wave moves towards the origin in the relative motion coordinates without returning.
This one-way traffic is reminiscent of what happens to matter or gravitational waves traveling towards the event horizon of a BH.
Moreover, when molecules interact with high partial-wave scatterings or dipole-dipole interactions, a potential barrier arises.
Near its maximum, the potential barrier is well approximated by an inverted parabola, similar to the barrier outside of the event horizon of a BH. 
Therefore, reactive molecules could serve as a natural quantum simulator of relevant BH physics outside the event horizon. %, as shown in Fig.\ref{Fig1}.  
In Fig.\ref{Fig1}, we show that the scattering of reactive molecules exhibits similar behavior as a scalar field scatters with the potential barrier outside the event horizon of a black hole.

In the literature, the unit reaction of molecules has been described as a ``black hole'' boundary condition based on the observation that matter waves go in without returning 
~\cite{BHboundary,BHboundary2}. However, there has been no attempt to formally connect reactive molecules to BH physics in a quantitative means. As we will show, the loss rate of reactive molecules exhibits a thermal-like distribution near the maximum of the potential barrier, in the same manner as a scalar field scatters with a BH. 
Thermal-like scatterings and connections to BH physics are hence readily accessible in current experiments. 
It is also worth mentioning that reactive molecules are highly controllable compared to other systems like quantum Hall states in two-dimensional electron gases. 
The potential barrier can be easily tuned by changing the angular momentum quantum number or
the strength of dipole-dipole interaction by varying the electric field strength. As such, the effective temperature in the thermal-like scattering is highly tunable.

\section{BH and IHO}
The spacetime metric of a Schwarzschild BH is written as 
\begin{align}
ds^{2}=f(r)d(ct)^{2}-f(r)^{-1}dr^{2}-r^{2}d\theta^{2}-r^{2}\sin^{2}\theta d\varphi^{2},
\end{align}
where $f(r)\equiv\left(1-r_{s}/r\right)$, and $r_{s}=2GM/c^{2}$ denotes the Schwarzschild radius with $G$ being the gravitational constant, $M$ the BH mass and $c$ the speed of light. The event horizon is located where $f(r)=0$, i.e., $r=r_{s}$. 
In the so-called ``tortoise'' coordinate $r_{*}$, $r_{*}=r+r_{s}\ln\left(r/r_{s}-1\right)$, the event horizon is moved to $r_{*}=-\infty$. A scalar field $\Phi(t,r,\theta,\varphi)$ interacting with a Schwarzschild BH satisfies 
\begin{align}
\label{BHradial}
\left(-\frac{d^{2}}{d r^{2}_{*}}+V_{\rm BH}(\ell,r_{*})\right)\psi_{\lambda}(r_{*})=\frac{\lambda^{2}}{c^{2}}\psi_{\lambda}(r_{*}),
\end{align}
where we have defined the single mode field with angular momentum $\ell$ as $\psi_{\lambda}(r_{*})=e^{i\lambda t}r\Phi(t,r_{*},\theta,\varphi)/Y^{m}_{\ell}(\theta,\varphi)$ with $\lambda$ being the frequency and $Y^{m}_{\ell}(\theta,\varphi)$ the spherical harmonic function. Eq.(\ref{BHradial}) is reminiscent of a stationary Schr\"odinger equation with potential $V_{\rm BH}(\ell,r_{*})$,
\begin{align}
\label{BHpotental}
V_{\rm BH}(\ell,r_{*})=\left(1-\frac{r_{s}}{r(r_{*})}\right)\left(\frac{r_{s}}{r(r_{*})^{3}}+\frac{\ell(\ell+1)}{r(r_{*})^{2}}\right).
\end{align} 
As shown in Fig.~\ref{Fig1}(a), $V_{\rm BH}(\ell,r_{*})$ has a potential barrier outside of the event horizon. Near the maximum of the potential, it can be approximated by the IHO. The larger is $\ell$, the better is the approximation.

Eq.(\ref{BHradial}) can be regarded as a Schr\"odinger equation of an IHO, and the Hamiltonian is written as
\begin{align}
\hat{H}=-\frac{1}{2}\frac{d^{2}}{dr_*^{2}}-\frac1{2}\omega^2(r_*-r_*^{\rm max})^2+\frac1{2}V_{\rm BH}(r_*^{\rm max}),
\end{align}
where $\omega=\sqrt{-\frac1{2}\partial^2_{r_*}V_{\rm BH}(\ell,r_*)|_{r^{\rm max}_*}}$ and $r_{*}^{\rm max}$ is defined via $\partial_{r_*}V_{\rm BH}(\ell,r_*)|_{r^{\rm max}_*}=0$. The two linear independent solutions to the Schr\"odinger equation $\hat{H}\psi_{E}(r_{*})=E\psi_{E}(r_{*})$ are $D_{iE/\omega-1/2}\left(i\sqrt{2i\omega }r_{*}\right)$ and $D_{-iE/\omega-1/2}\left(\sqrt{2i\omega }r_{*}\right)$ with $D_{a}(x)$ representing the parabolic cylinder function. Here, both $\omega$ and $E$ have the dimension $L^{-2}$. The energy and position have been measured from the top of IHO, i.e., $E\rightarrow E-V_{\rm BH*}/2$, $r_*\to r_*-r_*^{\rm max}$. 
By analyzing the asymptotic behavior of $\psi_E(r_*)$ in the limit of $r_{*}\to\pm\infty$, we could extract the scattering matrix 
\begin{align}
\label{IHO_smatrix}
S=\frac{\Gamma\left(\frac{1}{2}-\frac{iE}{\omega}\right)}{\sqrt{2\pi}}\left(\begin{array}{cc} -ie^{-\frac{\pi E}{2\omega}}& e^{\frac{\pi E}{2\omega}} \\ e^{\frac{\pi E}{2\omega}}& -ie^{-\frac{\pi E}{2\omega}}\end{array}\right),
\end{align}
where $\Gamma(*)$ denotes the Gamma function. Therefore, the reflection and transmission rate can be written as
\begin{align}
\label{IHO_reflection}
|r|^2_{\text{IHO}}&=\frac{1}{e^{\frac{2\pi E}{\omega}}+1};\quad |t|^2_{\text{IHO}}=\frac{1}{e^{-\frac{2\pi E}{\omega}}+1},
\end{align}
respectively, which follows a thermal-like distribution. The effective temperature 
$T_{\text{eff}}$ is thus determined by the IHO frequency $\omega$, which is related to the Schwarzschild radius. Note that it is a plus sign in the denominator, different from the minus sign in the expression for the Hawking radiation. As such, the thermal-like tunneling directly unfolds the mass of BH. 
When $\ell=0$, the effective temperature is written as
\begin{equation}
T_{\text{eff}}=\frac{27}{1024\sqrt{2}}\frac{\hbar^2c^4}{\pi k_B\tilde{\mu} M^2G^2 }.\label{BHtem}
\end{equation}
Since $k_BT_{\text{eff}}$ determined from Eq.(\ref{IHO_reflection}) has the dimension $L^{-2}$, the same as $\omega$, we have added in Eq.(\ref{BHtem}) an extra factor, $\hbar^2/\tilde{\mu}$, where $\tilde{\mu}$ is an arbitrary mass scale, to ensure that $T_{\text{eff}}$ has the same dimension as the temperature. We would like to point out that the thermal behavior of the scattering rate in Eq.(\ref{IHO_reflection}) is not due to Hawking radiation. The former originate from the elastic scattering at the top of potential barrier, while the later is due to quantum fluctuation near the event horizon.

For another $\ell$, the corresponding $T_{\text{eff}}$ can be obtained in the same way. It should be noted that there is a difference between $V_{\rm BH*}$ and $V_{\rm BH}(r_*^{\rm max})$, a constant energy shift from the top of the IHO to the maximum of the realistic potential. Such a shift exists when applying IHO as an approximation for a generic potential barrier, such as the P\"oschl-Teller potential that has analytical solutions~\cite{IHOPT}. With increasing $\ell$, the IHO approximation becomes better in the sense that it describes the potential barrier in a wider range of energy and the percentage difference between $V_{\rm BH*}$ and $V_{\rm BH}(r_*^{\rm max})$ decreases. 

In addition, the Rindler Hamiltonian that describes a reference frame moving with a constant acceleration and the resultant Unruh radiation turns out to be an IHO \cite{IHOPT}. Near the event horizon, the description of the surface gravity that produces the Hawking radiation also reduces to an IHO (Appendix \ref{appdenA}). Therefore, IHO plays a critical role underlying Hawking-Unruh radiation.

\section{Reactive molecules}
In the absence of an external electric field, two molecules interact via the van der Waals potential at large distance and the Hamiltonian of the relative motion is written as
\begin{align}
\label{vanderWaals}
\left[\frac{d^{2}}{dr^{2}}-\frac{\ell(\ell+1)}{r^{2}}+\frac{\beta_{6}^{4}}{r^{6}}+\frac{2\mu\epsilon}{\hbar^{2}}\right]u_{\ell}(r)=0,
\end{align}
where $\mu$ is the reduced mass, $\beta_{6}$ is the characteristic length of the van der Waals potential, and $u_{\ell}(r)=r\psi_{\ell}(r)$ with $\psi_{\ell}(r)$ being the radial wave function of $\ell$-th partial wave. The analytical solutions have been obtained by the quantum defect theory (QDT)~\cite{QDT1,QDT2}. The van der Waals potential and the centrifugal potential lead to an effective potential that has a maximum. As illustrated in Fig.~\ref{Fig1}(b),
near the potential maximum, the effective potential can be expanded as
\begin{align}
V_{\rm vdW}(r)\approx&-\frac{1}{2}\mu\omega^{2}\left(r-\frac{3^{1/4}\beta_{6}}{(\ell(\ell+1))^{1/4}}\right)^{2}+V_{\max},
\end{align}
where $\omega=2\ell(\ell+1)\hbar/(\sqrt{3}\mu \beta_{6}^{2})$, and the maximum of the potential is $V_{\max}=\hbar^2(\ell(\ell+1))^{3/2}/(3\sqrt{3}\mu \beta_{6}^{2})$. As such, similar to scatterings of a scalar field by BH, 
the high partial wave scattering of molecules can also be approximated by the IHO. The reaction with unit probability at short distance plays the role of an event horizon. Specifically, the asymptotic wave function at the short-range takes the following form
\begin{align}
\label{shortasymp}
u_{\ell}(r\to0)\propto&\frac{r^{3/2}}{\beta_{6}}
\exp[i\left(\frac{\beta^{2}_{6}}{2r^{2}}-\frac{\nu_{0}\pi}{2}-\frac{\pi}{4}\right)],
\end{align}
where $\nu_{0}=(2\ell+1)/4$. The absence of $\exp[-i\beta^{2}_{6}/(2r^{2})]$ in the wave function signifies unit probability of reaction such that there is no outgoing flux. 
As such, similar to the previously discussed BH physics, whereas we were considering a quantum tunneling problem, the transmission and reflection rate become thermal-like near the maximum of the potential barrier, 
\begin{equation}
T_{\text{eff}}=\frac{\ell(\ell+1)}{\sqrt{3}\pi}\frac{\hbar^{2}}{ k_B \mu \beta_6^2 }\label{VdWtem}.
\end{equation}
Comparing Eq.(\ref{BHtem}) and Eq.(\ref{VdWtem}), we see that, if identifying $\mu$ and $\tilde{\mu}$, $\beta_6$ plays the role of the mass of a BH, i.e., $\beta_6\sim M G/c^2$.

\begin{figure}
\centering
\includegraphics[width=0.45\textwidth]{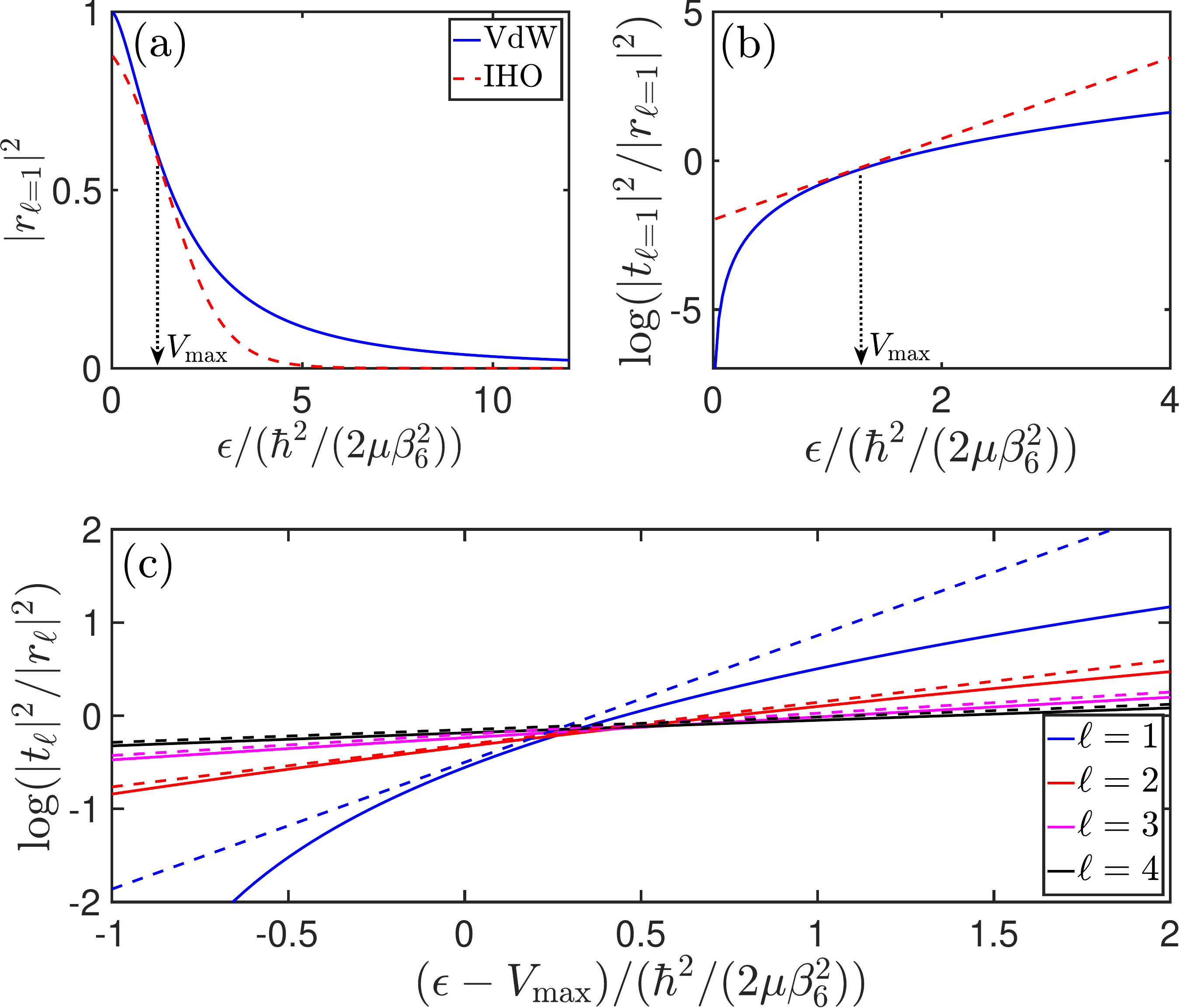}
\caption{The reflection and transmission rate for high partial-wave scatterings at zero electric field. (a) and (b) depict the reflection rate $|r|^2$ and $\log(|t|^2/|r|^2)$ as a function of the energy for $p$-wave scattering. (c): $\log(|t|^2/|r|^2)$ for various high partial-wave scatterings. 
In (a,b,c), solid curves are results from the quantum defect theory. Dashed curves are results from the approximation using IHO, whose frequency is determined by the potential near the maximum of the potential barrier. With increasing $\ell$, the approximation using IHO covers a broader range of energy.\label{Fig2}}
\end{figure}

The comparison between the exact results of the van der Waals potential and the result of IHO is shown in Fig.~\ref{Fig2}.  
Near the $V_{\max}$, an IHO well reproduces the result of the van der Waals potential. The slope of $\log (|t|^2/|r|^2)$ is given by the frequency of the IHO near the maximum of the barrier, as shown in Fig.~\ref{Fig2}(a,b).
Similar to BH scattering, there is a small difference between $V^*_{\max}$ and $V_{\max}$. We find $|V^*_{\max}-V_{\max}|/V_{\max}\approx 34\%$ for $\ell=1$.
We also find that with increasing $\ell$, the IHO approximation works well in a larger energy window near, as shown in Fig.~\ref{Fig2}(c). Meanwhile, $|V^*_{\max}-V_{\max}|/V_{\max}$ decreases. For $\ell=4$, $|V^*_{\max}-V_{\max}|/V_{\max}$ is readily as small as 3\%. Therefore, such thermal-like transmission and reflection become more evident in higher partial wave scatterings.

\begin{figure}
\centering
\includegraphics[width=0.4\textwidth]{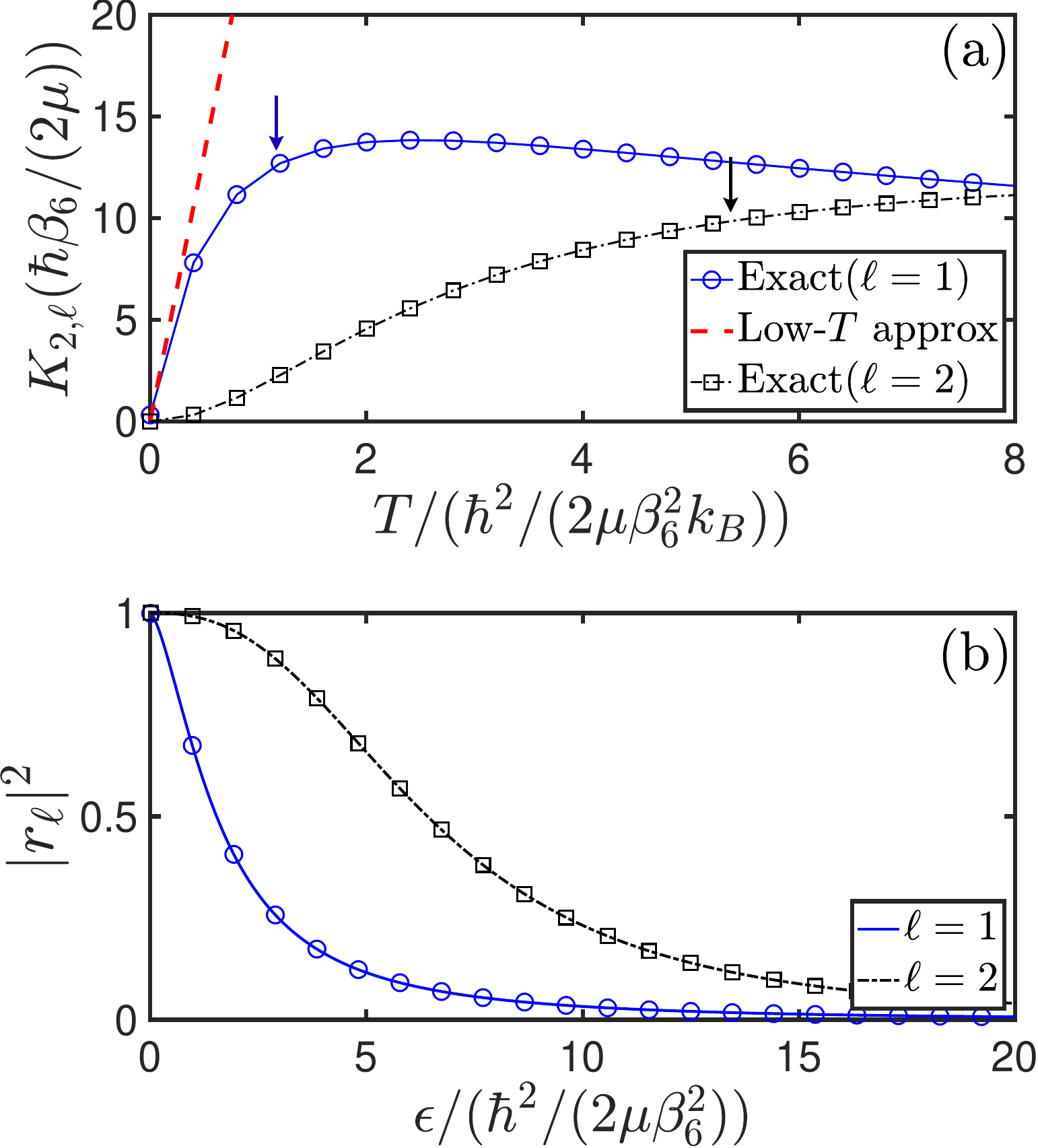}
\caption{(a) Two-body loss rates $K_{2,\ell}$ of $p$ (circle) and $d$-wave (square) scatterings. The blue and black arrows indicate the characteristic temperature that corresponds to the maximum of the potential barrier of $p$ and $d$-wave scattering, respectively. The Red dashed line depicts the low-temperature approximation of the $p$ wave scattering. (b) Markers are the results from an inverse Laplace transform of (a). Twenty temperature points in the $T$-axis have been used. It recovers the transmission rate as a function of the energy (solid and dash-dotted curves). 
\label{Fig3}}
\end{figure}

Whereas the energy-dependent transmission or reflection rate readily unfolds thermal-like scatterings in theory, what can be directly measured in experiments is the temperature-dependent two-body loss rate ${\cal K}_{\ell}^{\rm inel}(T)$. It is related to the transmission rate by a thermal average,
\begin{align}
{\cal K}_{\ell}^{\rm inel}(T)=(2\ell+1)\frac{4\pi\hbar^2}{(2\mu)^{3/2}}\frac{\int_{0}^{\infty}e^{-\frac{\epsilon}{k_{B}T}}|t|^2d\epsilon}{\int_{0}^{\infty}\sqrt{\epsilon}e^{-\frac{\epsilon}{k_{B}T}}d{ \epsilon}}. \label{thermalave}
\end{align}
 When the temperature is much smaller than the maximum of the barrier, we find that 
the loss rate is a constant and linearly dependent on $T$ for the $s$ and $p$-wave scatterings, respectively, i.e., ${\cal K}_{\ell=0}^{\rm inel}\approx4h\bar{a}/\mu$ and ${\cal K}_{\ell=1}^{\rm inel}\approx1512.58\bar{a}^{3}k_{B}T/h$ with $\bar{a}=2\pi\beta_{6}/\Gamma(1/4)^{2}$ and $h$ being the Planck constant, which are consistent with the results previously obtained in Ref.~\cite{Yejun0,Paul,Paul2,He2020}, as shown in Fig.~\ref{Fig3}(a).

Here, we are interested in the high-temperature regime of the order of $\mu$K, which is comparable to the typical value of the potential barrier maximum from the van der Waals interaction and the centrifugal potential. 
The scatterings near the maximum of the barrier thus become relevant. In Fig.~\ref{Fig3}(a), we show the two-body loss rate as a function of temperature for $p$ and $d$-wave 
scatterings. Though this thermal average convolutes the previously discussed thermal-like quantum tunnelings near the maximum of the potential barrier with scatterings at other energies,  Eq.(\ref{thermalave}) shows that such a thermal average is in fact a Laplace transform of the energy-dependent tunneling rate. An inverse Laplace transform thus could recover the energy-dependent reflection and transmission rate from the thermal averaged value. We have numerically confirmed that standard numerical techniques of the inverse Laplace transform are readily capable of recovering the thermal-like quantum tunneling from the thermally averaged decay rate. As shown in Fig. ~\ref{Fig3}(b), using Talbot's method, 
we could reproduce energy-dependent reflection and transmission rates from 20 data points of the thermal averaged decay rate around $T=4\hbar^2/(2\mu\beta_6^2k_B)$. This method works so well that the results by inverse Laplace transformation (markers) are indistinguishable from that given by QDT.

\begin{figure}
\centering
\includegraphics[width=0.4\textwidth]{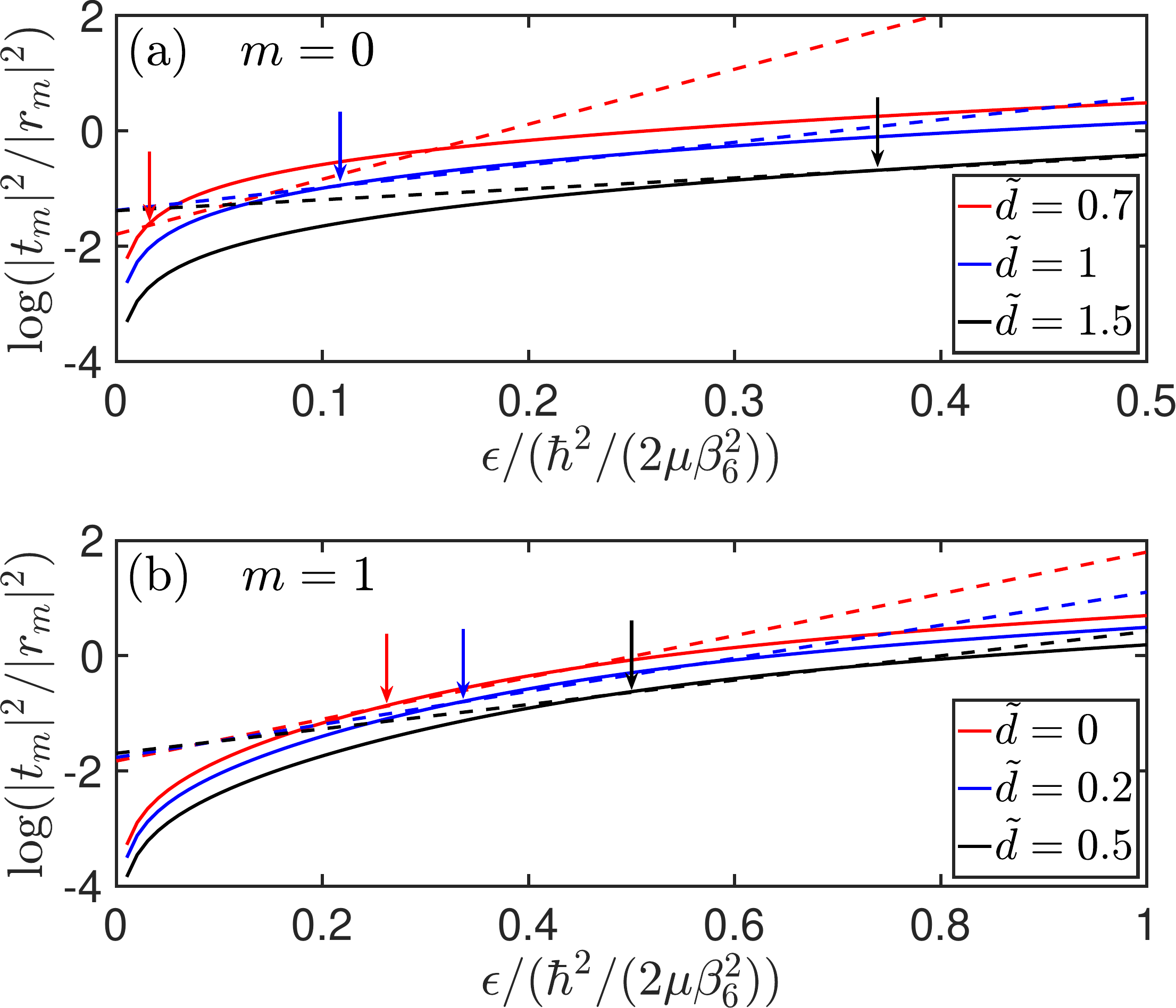}
\caption{ $\log(|t|^{2}/|r|^{2})$ as a function of the incident energy in the presence of an external electric field in 2D. (a) and (b) are results of $m=0$ and $m=1$, respectively. $\tilde{d}=2\mu d/(4\pi\hbar^{2}\epsilon_{0})$ denotes electric field induced dipole moment. Solid curves are results from the quantum defect theory. Dashed curves are the results from the approximation using IHO. Arrows indicate the energy of the potential maximum $V_{\max}$.\label{Fig4}}
\end{figure}

\section{Dipole-dipole interaction}
When an electric field is turned on, a dipole-dipole interaction between molecules is induced. 
Whereas such an interaction is anisotropic in three dimensions, to simplify discussions, we consider two dimensions and the electric field is perpendicular to the plane~\cite{2DYejun,2DYejun1}. As a result, an isotropic dipole-dipole repulsion creates a potential barrier even for $s$-wave scatterings. The dipole-dipole interaction depends on the electric field, providing another knob to control the thermal-like tunnelings.
The Schr\"odinger equation along the radial direction reads,
\begin{align}
\label{dipole}
\left(\frac{d^{2}}{d\rho^{2}}-\frac{m^{2}-1/4}{\rho^{2}}+\frac{\beta_{6}^{4}}{\rho^{6}}-\frac{2\mu d^{2}}{4\pi\hbar^{2}\varepsilon_{0}\rho^{3}}+\frac{2\mu \epsilon}{\hbar^{2}}\right)u_{m}(\rho)=0,
\end{align}
where $\varepsilon_{0}$ is vacuum permittivity, and $d$ is the induced electric dipole moment that depends on the electric field. $m$ is the quantum number of the angular momentum about the normal direction of the plane.
Since there is no simple analytical solution to Eq.(\ref{dipole}), we numerically solve it and extract the scattering properties.

Fig.~\ref{Fig4} shows the log of $|t_{m}|^{2}/|r_{m}|^{2}$ with a varying electric dipole moment for the $s$-wave and $p$-wave scattering. In most current experiments, the trapping potential height is around a van der Waals energy $\hbar^{2}/(2\mu\beta_{6}^{2})$. We thus consider incident energies smaller than $\hbar^{2}/(2\mu\beta_{6}^{2})$. By increasing the dipole moment, the IHO approximation works better and better. Even for the $s$-wave scattering, thermal scattering can be observed. In other words, the linear region near $V_{\rm max}$ becomes broader and broader by increasing the electric field, which is hopefully to be observed in current experiments.

Whereas we have been focusing on unit reaction probabilities at a short distance, it is worth considering smaller reaction probabilities. The reactive rate is characterized by a dimensionless ``quantum-defect'' parameter $0\leq y\leq1$~\cite{Paul}. $y=1$ and $y=0$ indicate that the molecule collision at a short range is complete lossy and elastic, respectively. Experiments have reported $y= 0.26, 1.0$ for RbCs, KRb~\cite{Gregory2019,Yejun3}. 
 Recent experiments have further shown that $y$ can be manipulated by the external magnetic or electric field, which offers an unprecedented means to 
 tune the boundary condition from a perfect event horizon to an imperfect one that partially or totally reflects the incident wave~\cite{Yejun,Yejun2,Ketterle}. In the latter case, matter waves bounce back and forth between the potential barrier and the imperfect event horizon. Again, near the maximum of the potential barrier, such a multiple scattering problem of IHO well captures the exact result. The thermal-like tunneling can therefore be extracted from the decay rate of any $y$ (Appendix \ref{appdenB}).

We have shown that the reactive molecules allow physicists to simulate scatterings between gravitational waves and BHs. The reactive molecules are also promising candidates for studying other profound features of BHs, such as quasi-normal modes and BH ringdown, if the time-dependence of the reflected waves can be probed in experiments. 
We hope that our work will stimulate more interest in studying the BH physics using AMO systems.

\begin{acknowledgments}
We are grateful to Jun Ye and Paul Julienne for their helpful discussions. R.Z. is supported by NSFC (Grant No.12174300) and the National Key R$\&$D Program of China (Grant No. 2018YFA0307601). Q.Z. and C.L. are supported by National Science Foundation (NSF) through Grant No. PHY-2110614.
\end{acknowledgments}

\appendix

\section{Hawking radiation and IHO}
\label{appdenA}
In this section, we show that the equation of motion of a scalar field near the event horizon of Schwarzschild black hole (BH) is of the same form as the stationary Schr\"odinger equation of the inverted harmonic oscillator (IHO). Near the event horizon, the Schwarzschild metric can be reduced to the Rindler metric which describes the spacetime of a uniformly accelerating system. To this end, we define 
\begin{align}
\rho&=\int_{r_{s}}^{r}\left(1-\frac{r_{s}}{r'}\right)^{-1/2}dr'\nonumber\\
&=\sqrt{r(r-r_{s})}+r_{s}\tanh^{-1}\left(\sqrt{1-\frac{r_{s}}{r}}\right).
\end{align}
Near the event horizon, $\rho$ can be approximated by
\begin{align}
\rho\approx2\sqrt{r_{s}(r-r_{s})},
\end{align}
and the  Schwarzschild metric can be written as
\begin{align}
ds^{2}\approx\rho^{2}\left(\frac{cdt}{2r_{s}}\right)^{2}-d\rho^{2}-\left(\frac{\rho^{2}}{4r_{s}}+r_{s}\right)^{2}(d\theta^{2}+\sin^{2}\theta d\varphi^{2}).
\end{align}
Due to the rotational symmetry, we could only consider the radial part. By defining a dimensionless time $\tau=ct/(2r_{s})$, we have the Rindler metric in the $1+1$ spacetime,
\begin{align}
ds^{2}=\rho^{2}d\tau^{2}-d\rho^{2},
\end{align}
which can also be written into a conformally flat form by choosing a new spatial coordinate $\xi=\kappa^{-1}\ln(\kappa\rho)$ and rescaling $\tau\rightarrow \tau/\kappa$. Then we have the following metric
\begin{align}
ds^{2}=e^{2\kappa\xi}(d\tau^{2}-d\xi^{2}).
\end{align}
Then equation of motion of scalar field near the event horizon can be written as
\begin{align}\label{tx}
\left(\frac{\partial^{2}}{\partial\tau^{2}}-\frac{\partial^{2}}{\partial\xi^{2}}\right)\phi(\tau,\xi)=0.
\end{align}
The eigen modes are
\begin{align}
\label{em}
\phi_{\pm}(\tau,\xi)=e^{\pm i\Omega(\xi\mp\tau)},
\end{align}
which satisfy $i\partial_\tau\phi_{\pm}=\Omega \phi_\pm$ and
can be understood as the right/left-moving modes. 
If we define 
\begin{align}\label{uvxt}
u=-\frac{e^{\kappa(\xi-\tau)}}{\kappa};\quad v=\frac{e^{\kappa(\xi+\tau)}}{\kappa},\end{align}
and the scaling operators
\begin{align}
\label{scalingBvu}
\hat{\cal S}_{v}=-iv\partial_{v};\quad \hat{\cal S}_{u}=iu\partial_{u},
\end{align}
Eq.(\ref{tx}) becomes
\begin{align}
\hat{\cal S}_{u}\hat{\cal S}_{v}\phi(u,v)=0. 
\end{align}
Noticing that $\partial_\tau=i\kappa( \hat{\cal S}_{u}+ \hat{\cal S}_{v})$,
the two independent solutions $\phi_1(u,v)\equiv\phi_1(u)$ and $\phi_2(u,v)\equiv\phi_2(v)$ thus satisfy
\begin{align}\label{uv}
\hat{\cal S}_{u}\phi_1(u)=-\frac{\Omega}{\kappa} \phi_1(u);\quad \hat{\cal S}_{v}\phi_2(v)=-\frac{\Omega}{\kappa} \phi_2(v),
\end{align}
and
\begin{align}
\label{eigenmode}
\phi_1(u,v)=(-\kappa u)^{ i\Omega/\kappa},\,\,\,\,\, \phi_2(u,v)=(\kappa v)^{- i\Omega/\kappa},
\end{align}
which can also be obtained from Eq.(\ref{em}) using Eq.(\ref{uvxt}).
Eq.~(\ref{eigenmode}) as eigenfunctions of $\partial_\tau$ 
and the scaling operators $\hat{\cal S}_{u/v}$ 
underlie the Hawking-Unruh radiation
~\cite{Hawkingradiation,Unruh0}.

On the other hand, The Hamiltonian of the inverted harmonic oscillator is written as
\begin{align}
\hat{H}_{\rm IHO}=\frac{\hat{p}^{2}}{2m}-\frac{1}{2}m\omega^{2}\hat{x}^{2}.
\end{align}
We define a new set of variables 
\begin{align}
\hat{u}^{\pm}=\frac{\hat{p}\pm m\omega \hat{x}}{\sqrt{2m\omega}},
\end{align}
and it can be check that $[\hat{u}^{+},\hat{u}^{-}]=i\hbar$. Then, the IHO Hamiltonian can be rewritten as
\begin{align}
\label{IHOplus}
\hat{H}_{\rm IHO}=\frac{\omega}{2}\left(\hat{u}^{+}\hat{u}^{-}+\hat{u}^{-}\hat{u}^{+}\right).
\end{align}
Now we consider the solutions to the Schr\"odinger equation in two distinct representations: 
\begin{itemize}
\item In the $u^{+}$-space, we have 
\begin{align}
\label{IHOplus}
\hat{H}_{\rm IHO}=\frac{\omega}{2}\left(2\hat{u}^{+}\hat{u}^{-}-i\hbar\right)=-i\hbar\omega\left(u^{+}\partial_{u^{+}}+\frac{1}{2}\right),
\end{align}
and the solution to the Schr\"odinger equation $\hat{H}_{\rm IHO}\psi(u^{+})=E\psi(u^{+})$ reads $\psi(u^{+})=(\pm u^{+})^{\frac{iE}{\hbar\omega}-\frac{1}{2}}$.
\item In the $u^{-}$-space, we have
\begin{align}
\label{IHOminus}
\hat{H}_{\rm IHO}=\frac{\omega}{2}\left(2\hat{u}^{-}\hat{u}^{+}+i\hbar\right)=i\hbar\omega\left(u^{-}\partial_{u^{-}}+\frac{1}{2}\right),
\end{align}
and the solution to the Schr\"odinger equation $\hat{H}_{\rm IHO}\psi(u^{-})=E\psi(u^{-})$ reads $\psi(u^{+})=(\pm u^{-})^{-\frac{iE}{\hbar\omega}-\frac{1}{2}}$.
\end{itemize}
Comparing Eq.(\ref{uv}) with Eq.(\ref{IHOplus}), (\ref{IHOminus}), we find that near the event horizon, the dynamic of a scalar field, up to a constant energy shift, can be reduced to that of IHO.

\section{Imperfect event horizon}
\label{appdenB}
\begin{figure*}
\centering
\includegraphics[width=0.6\textwidth]{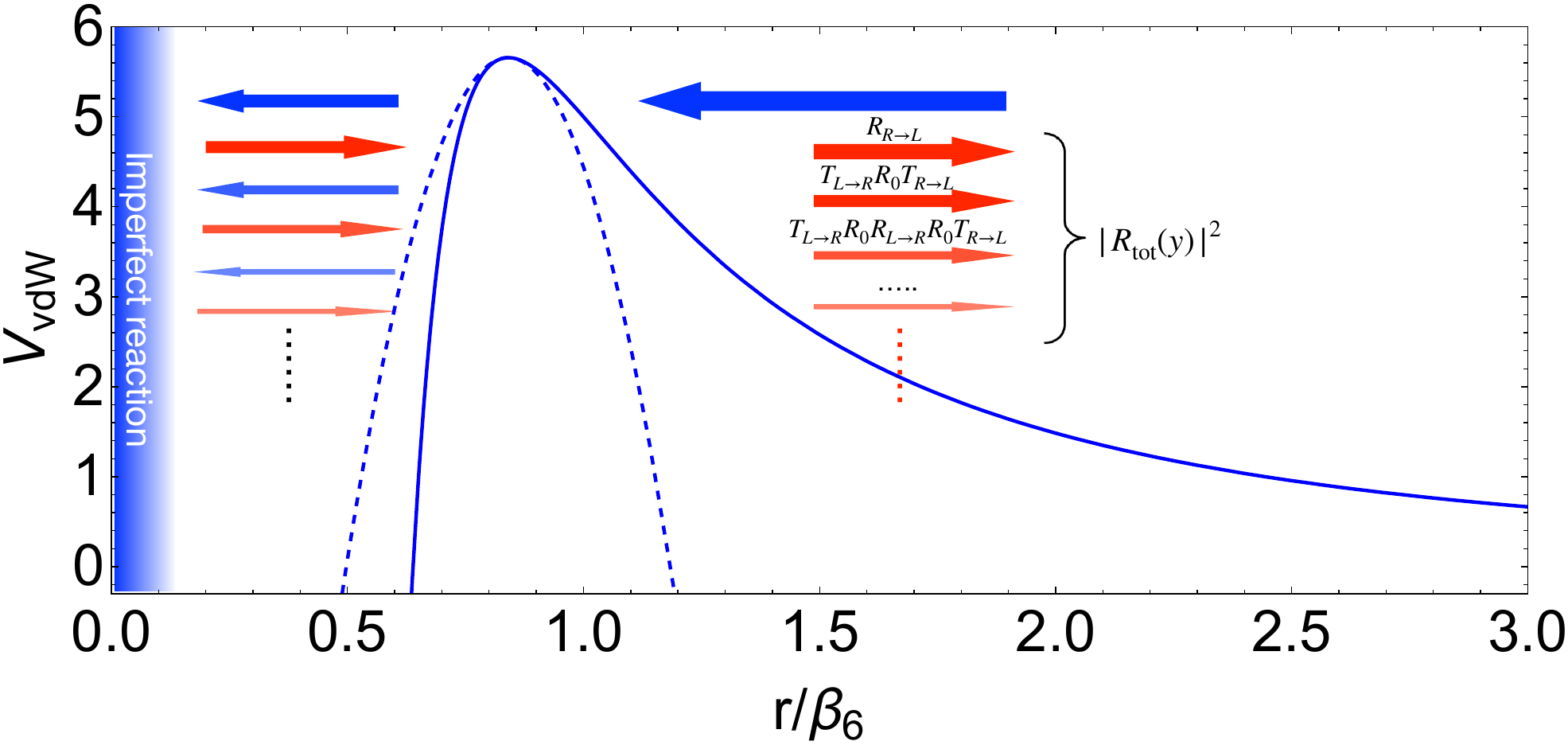}
\caption{A schematic of multiple scatterings caused by an imperfect event horizon. $R_{R(L)\to L(R)}$ and $T_{R(L)\to L(R)}$ indicate the reflection and transmission amplitude for the ``right(left) to left(right)'' scattering. $R_{\rm tot}(y)$ denotes the total reflection amplitude with generic ``quantum parameter'' $y$. It can be obtained by summing over such an infinite series of scatterings. Blue and red arrows denote the left and right 
moving waves. The solid and dashed  curves depict the interaction between two molecules with high partial-wave scatterings (or dipole-dipole interactions) and the IHO approximation, respectively.\label{Supfig0}}
\end{figure*}

The reflection and transmission with a generic ``quantum defect'' parameter $y$ can be obtained by studying  multiple scatterings between an imperfect event horizon and potential barrier, as shown in Fig.~\ref{Supfig0}.
The imperfect event horizon partially reflects the incoming waves and thus corresponds to a reaction rate less than unity.
According to the quantum defect theory (QDT), the short range asymptotic behavior of the radial wave function is written as
\begin{align}
\label{asympu}
u_{\ell}(r\to0)\propto&\frac{r^{3/2}}{\beta_{6}}\left[e^{i\left(\frac{1}{2}\left(\frac{\beta_{6}}{r}\right)^{2}-\frac{\nu_{0}\pi}{2}-\frac{\pi}{4}\right)}\right.\nonumber\\
&\left.-\frac{1-\frac{1}{iK_{\ell}^{0}}}{1+\frac{1}{iK_{\ell}^{0}}}e^{-i\left(\frac{1}{2}\left(\frac{\beta_{6}}{r}\right)^{2}-\frac{\nu_{0}\pi}{2}-\frac{\pi}{4}\right)}\right]
\end{align}
where $K_{\ell}^{0}$ is the $\ell$-th wave K-matrix \cite{QDT2}. We define the ``quantum-defect'' parameter $y$ in the following way
\begin{align}
\frac{1-\frac{1}{iK_{\ell}^{0}}}{1+\frac{1}{iK_{\ell}^{0}}}=\frac{1-y}{1+y}e^{-2i\eta_{\ell}},
\end{align}
where $\eta_{\ell}$ is the scattering phase shift. For the case that $y=1$, $K_{\ell}^{0}=-i$, and the second term in Eq.(\ref{asympu}) vanishes, which corresponds to complete absorption. For the case that $y=0$, $K_{\ell}=-\frac{1}{\tan\eta_{\ell}}$, which corresponds to a complete reflection,  i.e., a perfect event horizon. For the generic case, the K-matrix is $y$-dependent via,
\begin{align}
K_{\ell}^{0}(y)=-\frac{1}{\tan\eta_{\ell}}+y\frac{1+\tan^2\eta_{\ell}}{(y+i\tan\eta_{\ell})\tan\eta_{\ell}}.
\end{align}
The long-range asymptotic behavior of the radial wave function is written as
\begin{align}
&u_{\ell}(r\to\infty)\propto\nonumber\\
&\left(K_{\ell}^{0}(y)Z_{gg}-Z_{fg}+i\left(K_{\ell}^{0}(y)Z_{gf}-Z_{ff}\right)\right)e^{i\left(kr-\frac{\ell \pi}{2}\right)}\nonumber\\
&+\left(K_{\ell}^{0}(y)Z_{gg}-Z_{fg}-i\left(K_{\ell}^{0}(y)Z_{gf}-Z_{ff}\right)\right)e^{-i\left(kr-\frac{\ell \pi}{2}\right)},
\end{align}
where $Z_{gg},Z_{gf},Z_{fg},Z_{ff}$ are the element of the Z-matrix which is defined in Ref. \cite{QDT2}. Then we find the reflection amplitude is written as
\begin{align}
R(y)=\frac{K_{\ell}^{0}(y)Z_{gg}-Z_{fg}+i\left(K_{\ell}^{0}(y)Z_{gf}-Z_{ff}\right)}{K_{\ell}^{0}(y)Z_{gg}-Z_{fg}-i\left(K_{\ell}^{0}(y)Z_{gf}-Z_{ff}\right)}.
\end{align}
We would like to point out that this result can be explained by taking into account an infinite series of bounces between the potential barrier and the imperfect event horizon. To this end, we distinguish the ``right-to-left scattering'' and ``left-to-right scattering'' through the potential barrier.
\begin{itemize}
\item Right to left scattering; In such case, the short range boundary condition is written as
\begin{align}
u_{R\to L}(r\to0)\propto&\frac{r^{3/2}}{\beta_{6}}e^{i\left[\frac{1}{2}\left(\frac{\beta_{6}}{r}\right)^{2}-\frac{\nu_{0}\pi}{2}-\frac{\pi}{4}\right]}
\end{align}
which means $K_{\ell}^{0}=-i$, and then we have
\begin{align}
&u_{R\to L}(r\to\infty)\propto\nonumber\\
&\left(Z_{gf}-Z_{fg}-iZ_{ff}-iZ_{gg}\right)e^{i\left(kr-\frac{\ell \pi}{2}\right)}\nonumber\\
&+\left(iZ_{ff}-iZ_{gg}-Z_{gf}-Z_{fg}\right)e^{-i\left(kr-\frac{\ell \pi}{2}\right)},
\end{align}
and the corresponding reflection and transmission amplitude are
\begin{align}
\label{RL}
R_{R\to L}&=\frac{Z_{gf}-Z_{fg}-i(Z_{ff}+Z_{gg})}{i(Z_{ff}-Z_{gg})-(Z_{fg}+Z_{gf})};\\
 T_{R\to L}&=\frac{2\sqrt{2}}{i(Z_{ff}-Z_{gg})-(Z_{gf}+Z_{fg})},
\end{align}
respectively.
Using the fact that $Z_{ff}Z_{gg}-Z_{fg}Z_{gf}=-2$, it is readily to verify $|R_{R\to L}|^{2}+|T_{R\to L}|^{2}=1$.

\item Left to right scattering; In such case, the long range boundary condition is written as
\begin{align}
&u_{L\to R}(r\to\infty)\propto\nonumber\\
&\left(\left(K_{\ell}^{0}Z_{gg}-Z_{fg}\right)+i\left(K_{\ell}^{0}Z_{gf}-Z_{ff}\right)\right)e^{i\left(kr-\frac{\ell \pi}{2}\right)}\nonumber\\
&+\left(\left(K_{\ell}^{0}Z_{gg}-Z_{fg}\right)-i\left(K_{\ell}^{0}Z_{gf}-Z_{ff}\right)\right)e^{-i\left(kr-\frac{\ell \pi}{2}\right)}.
\end{align}
Since the second term is required to vanish, we have 
\begin{align}
\left(K_{\ell}^{0}Z_{gg}-Z_{fg}\right)-i\left(K_{\ell}^{0}Z_{gf}-Z_{ff}\right)=0,
\end{align}
i.e.,
\begin{align}
K_{\ell}^{0}=\frac{Z_{fg}-iZ_{ff}}{Z_{gg}-iZ_{gf}}.
\end{align}
Then the short range boundary condition can be written as
\begin{align}
&u_{L\to R}(r\to0)\propto\nonumber\\
&\frac{r^{3/2}}{\beta_{6}}\left[\frac{Z_{gg}+Z_{ff}+i(Z_{fg}-Z_{gf})}{Z_{gg}-iZ_{gf}}e^{i\left(\frac{1}{2}\left(\frac{\beta_{6}}{r}\right)^{2}-\frac{\nu_{0}\pi}{2}-\frac{\pi}{4}\right)}\right.\nonumber\\
&\left.+\frac{Z_{gg}-Z_{ff}-i(Z_{fg}+Z_{gf})}{Z_{gg}-iZ_{gf}}e^{-i\left(\frac{1}{2}\left(\frac{\beta_{6}}{r}\right)^{2}-\frac{\nu_{0}\pi}{2}-\frac{\pi}{4}\right)}\right],
\end{align}
and the corresponding reflection and transmission amplitude are written as
\begin{align}
\label{LR}
R_{L\to R}&=\frac{(Z_{fg}-Z_{gf})-i(Z_{gg}+Z_{ff})}{i(Z_{ff}-Z_{gg})-(Z_{fg}+Z_{gf})};\\
 T_{L\to R}&=\frac{\sqrt{2}(Z_{fg}Z_{gf}-Z_{ff}Z_{gg})}{i(Z_{ff}-Z_{gg})-(Z_{fg}+Z_{gf})},
\end{align}
respectively. It can also be checked that $|R_{L\to R}|^{2}+|T_{L\to R}|^{2}=1$.
\end{itemize}
We use $R_{0}=-\frac{1-y}{1+y}e^{2i\eta_{\ell}}$ to characterize the reflection amplitude near the event horizon, where $\eta_{\ell}$ is the phase shift of the elastic scattering. Then the total reflection amplitude can be written as
\begin{align}
\label{Rtot}
R_{\rm tot}(y)=&R_{R\to L}+T_{L\to R}R_{0}T_{R\to L}+T_{L\to R}R_{0}R_{L\to R}R_{0}T_{R\to L}\nonumber\\
&+T_{L\to R}R_{0}R_{L\to R}R_{0}R_{L\to R}R_{0}T_{R\to L}+...\nonumber\\
=&R_{R\to L}+T_{L\to R}\frac{R_{0}}{1-R_{0}R_{L\to R}}T_{R\to L}.
\end{align}
By substituting Eq.(\ref{RL}) and Eq.(\ref{LR}) into Eq.(\ref{Rtot}), we immediately find $R_{\rm tot}(y)=R(y)$. As a result, the thermal-like tunneling can be extracted from the decay rate of any $y$.

\begin{figure*}
\centering
\includegraphics[width=0.8\textwidth]{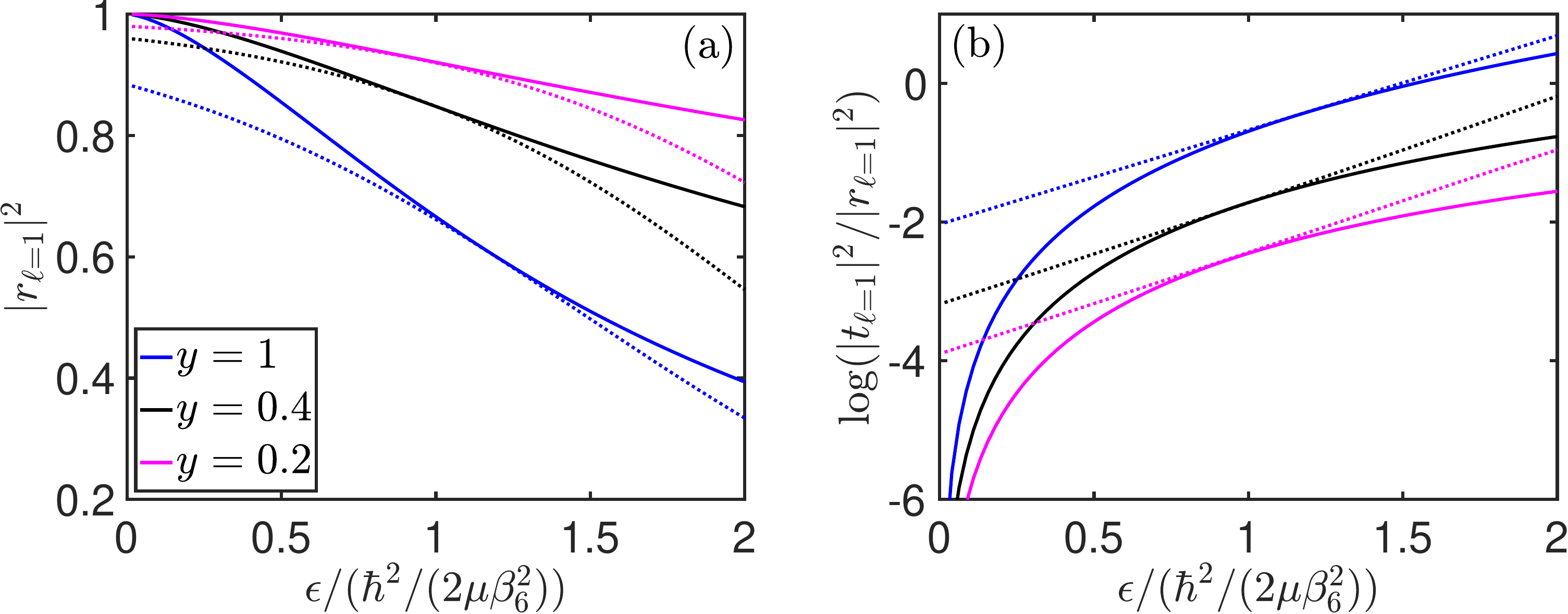}
\includegraphics[width=0.8\textwidth]{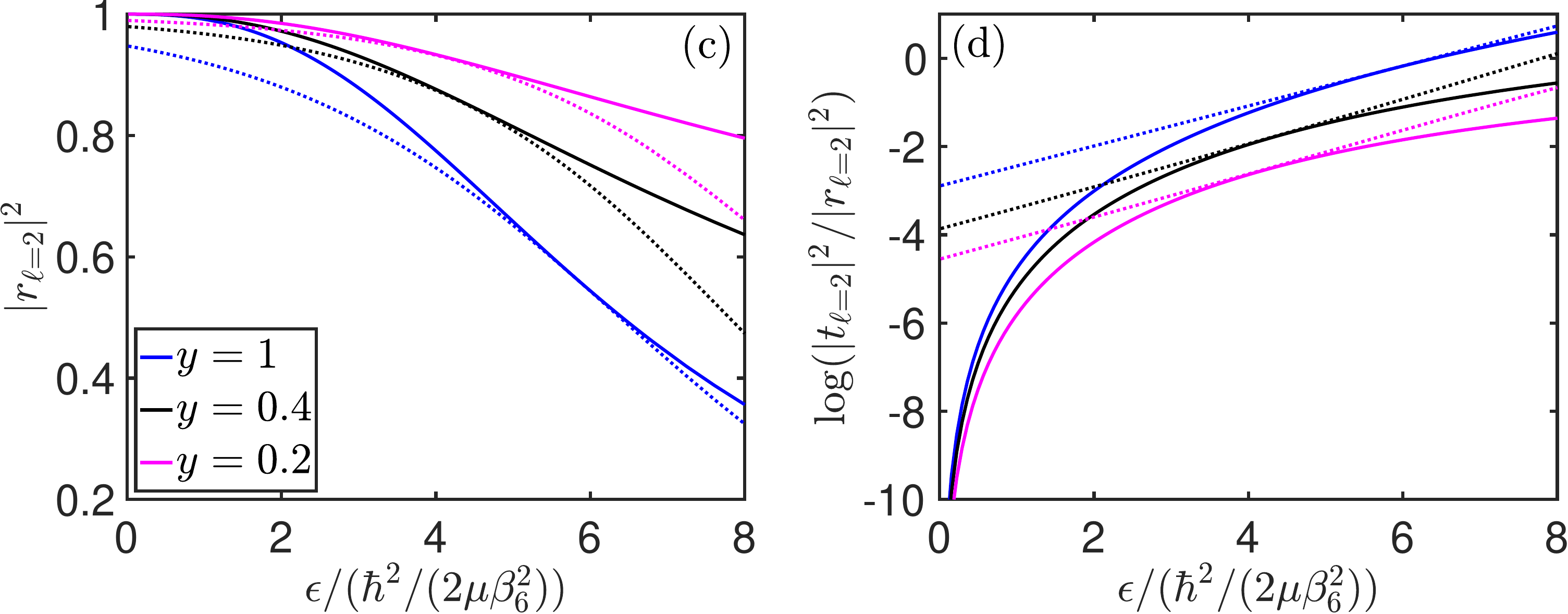}
\caption{Reflection rate $|r_{\ell}|^{2}$ and $\log(|t_{\ell}|^{2}/|r_{\ell}|^{2})$ of reactive molecule under imperfect absorbing boundary condition for $p$-wave (a,b) and $d$-wave (c,d) scattering.The solid curves depict the results of van der Waals potential and the dashed curves show the results of IHO.\label{Supfig1}}
\end{figure*}

In the same spirit, we can obtain the scattering amplitude of IHO by implementing an imperfect absorbing boundary condition at 
the imperfect event horizon. Considering the reflection symmetry of the IHO potential, we have 
\begin{align}
|r_{\rm IHO}|^{2}(y)=&\left|\frac{R-R_{0}(R^{2}-T^{2})}{1-R_{0}R}\right|^{2},
\end{align}
where $R=S_{11}$ and $T=S_{12}$ are the reflection and transmission amplitude of IHO, as shown in Eq.(5) of the main text. In Fig.~\ref{Supfig1}, we show the reflection rate of van der Waals potential under the partial absorbing boundary condition provided by an imperfect event horizon. It is clear that the IHO provides a good approximation for a generic $y$.

%\bibliographystyle{apstest}
%\bibliography{ref_BH.bib}

\end{document}